\documentclass[fleqn,10pt,twoside]{article}

\usepackage[english] {babel}

\usepackage {graphicx}
\usepackage {graphics}
\usepackage {floatflt}
\usepackage {latexsym}
\usepackage {caption3} 
\usepackage[labelsep=period]{caption}[2007/11/04 v.3.1e]
\usepackage {amssymb}

\def\noi{\noindent}

\renewcommand{\thesubsubsection}%
        {\arabic{section}.\arabic{subsection}.\arabic{subsubsection}.}

\newcommand{\heads}[2]{\markboth{\protect\small\it #1}{\protect\small\it #2}}

\newcommand{\Arthead}[5]{ \setcounter{page}{#4}\thispagestyle{empty}\noi
    \unitlength=1pt \begin{picture}(500,40)

        \put(0,58){\shortstack[l]{\small\it ISSN 0202-2893, Gravitation \& Cosmology,
                          #1, Vol.#2, No. #3, pp. #4--#5    
\footnotesize\copyright \ Pleiades Publishing Ltd., 2012.} }

    \end{picture}
	 }     		
\def\prepno#1#2
    {\thispagestyle{empty}
    \noi    \unitlength=1mm
    	\begin{picture}(178,10)
            \put(177,15){\llap{\bf #1}}
            \put(177,10){\llap{\small\rm #2}}
        \end{picture}
          }             

\newcommand{\Title}[1]{\noi {\uppercase{\Large #1}}     }
\newcommand{\Author}[2]{\noi{\large\bf #1}\\[2ex]\noindent{\it #2}   }

\newcommand{\Abstract}[1]{\vskip 2mm \begin{center}
        \parbox{16.4cm}{\small\noi #1} \end{center}\medskip}

\newcommand{\foom}[1]{\protect\footnotemark[#1]}

\newcommand{\email}[2]{\footnotetext[#1]{e-mail: #2}
		\addtocounter{footnote}{1}}

\heads{A.M.Baranov}
      {On an approach to constructing static ball models in General Relativity} %

\begin{document}
\twocolumn 
[
\Arthead{2012}{18}{3}{201}{203.}

\vspace{0.5cm}
\begin{center}
\Title{On an approach to constructing static ball models

\vspace{0.2cm}
  in General Relativity \foom1}
\end{center}

\begin{center}
\vspace{.5cm}
   \Author{A.M.Baranov\foom2}   
{\it Dept. of Physics, Siberian State Technological University,
82 Mira Av. Krasnoyarsk, 660049, Russia}

{\it Received October 13, 2011}
\end{center}

\Abstract
{{\bf Abstract} -- An approach to construction of static models is demonstrated for a fluid ball. Five examples are considered. Two of them are exact solutions of the Einstein equations; the other three are connected with the  Airy special functions, the hypergeometric functions and the Heun functions.}

{\bf DOI:} 10.1134/S020228931203324
\vspace{1.0cm}
] 

\footnotetext[1]{Talk given at the International Conference RUSGRAV-14, 
 June 27--July 4, 2011, Ulyanovsk, Russia.}

\email 2 {alex\_m\_bar@mail.ru }

\section{Introduction}

\medskip
One of the most important problems in general relativity is that of finding exact solutions of the gravitational equations. Unfortunately, this task is not easy. Another way is an analytic construction of solutions to the gravitational equations with a certain physical interpretation. 

Further on we take the metric in Bondi's form 

$$ds^2 = G(r)^2 dt^2 +2L(r) dtdr $$
$$ 
- r^2(d\theta^2+sin^2(\theta)d\varphi^2) 
\eqno{(1)} 
$$
\noindent
where $G^2, L$ are metric functions, $r$ is a radial variable and $\theta, \varphi$ are spherical angles; the speed of light and Newton's constant of gravity are put equal to unit.

\medskip
The gravitational field is described by the metric tensor $g_{ik},$ which can be found from Einstein's equations 
$$
R_{ik} - \frac{1}{2}Rg_{ik} = - \varkappa T_{ik},
\label{eq:2}
\eqno{(2)} 
$$
\noindent
where $i,j,k = 0, 1, 2, 3$; $R_{ik}$ is the Ricci tensor, $R$ is the scalar curvature of space-time; $\varkappa$ is Einstein's gravitational constant.

\medskip
The energy-momentum tensor (EMT) of Pascal's perfect fluid can be written as 
$$
T_{ik}  = \left( {p\left( r \right) + \mu \left( r \right)} \right) \cdot u_i u_k  - p\left( r \right)g_{ik},
\eqno{(3)} 
$$
\noindent
where $p(r)$ is the pressure, $\mu(r)$ is the mass density, $u_i$ is the 4-velocity. 

The gravitational equations in dimensionless variables can be reduced after elementary transformations to the form 
$$
\varepsilon (x) = 1 - \frac{\chi }{x} \cdot \int {\mu (x) \cdot x^2 dx};
\eqno{(4)} 
$$

$$
G^{\prime\prime}+\left( {\frac{{\varepsilon '}}{{2 \cdot \varepsilon }} - \frac{1}{x}} \right) G^\prime + \left( {\frac{{\varepsilon '}}{{2 \cdot x \cdot \varepsilon }} + \frac{{1 - \varepsilon }}{{x^2  \cdot \varepsilon }}} \right) G = 0;
\eqno{(5)} 
$$

$$
p^{\prime} = - \displaystyle{\frac{1}{{2 \cdot \varepsilon }} \cdot \left( {\chi x  p + \frac{{1 - \varepsilon }}{x}} \right) \cdot \left( {\mu  + p} \right)},
\eqno{(6)} 
$$
\noindent
where $x = r/R$ is the dimensionless radius;$\;0 \leq x \leq 1;\;$ differentiation in $x$ is denoted by a prime; $R$ is the radius of the astrophysical object; $\chi=\varkappa\cdot R^2;$ 

$$
\varepsilon (x) = \displaystyle\frac{G^2(x)}{L ^2(x)}.
\eqno{(7)} 
$$

\section{Reduction of the Einstein \\
equations}

Eq (5) can be reduced to an oscillatory-type equation 
$$
\displaystyle\frac{d^2 G}{d \zeta^2} + \Omega^2(\zeta(y)) G = 0,
\eqno{(8)} 
$$
\noindent
where 
$$
\Omega^2 = -d(\Phi/y)/dy,
\eqno{(9)} 
$$
\noindent
$y=x^2,$ $\zeta$ is a new variable: 
$$
d\zeta = y dy/(2 \sqrt{\varepsilon}).
\eqno{(10)} 
$$

Now we introduce $\varepsilon = 1-\Phi$ as in \cite{1}, where
$$
\Phi = (\chi/(2\sqrt{y})\int {\mu \sqrt{y}dy},
\eqno{(11)} 
$$
\noindent
where $\chi = \varkappa R^2,$ $\Phi$ is an analog of Newton's gravitational potential.

Further we present the function $\Omega^2$ in the form of a power series:
$$
\Omega^2(y) = \sum_{n=0}^{\infty} a_n y^n.
\eqno{(12)} 
$$

Now we can find both $\Phi$ and $\mu$ in the general case from (8) and (10) with help of (11)
$$
\Phi = (\mu_0/3) y - \sum_{n=0}^{\infty} \displaystyle\frac{a_n}{n+1} y^{n+2}; \\
\eqno{(13)} 
$$

$$
\mu/\mu_0 = 1 - (1/\mu_0)\sum_{n=0}^{\infty} \displaystyle\frac{2n+3}{n+1}a_n y^{n+1},
\eqno{(14)} 
$$
\noindent
where $\mu_0$ is the central mass density and the coefficients $a_n$ are constants to be found from boundary conditions.

\section{Construction of solutions  \\
to the gravitational equations}

Consider the power series (11). At first we will take all coefficients $a_n$ equal to zero. After that we will consider only $a_0 \neq 0,$ further $a_0 \neq 0,$ $a_1 \neq 0$ and so on. In each case we will construct the corresponding mass density and make an attempt to find the function $G$ from (7). Here we must remark that the metric functions $g_{00} = G^2$ and $g_{10} = g_{01} \equiv L = G /\sqrt{\varepsilon}.$ 

{\bf 1.} If we take all $a_n=0$, then $\Omega^2=0$ and Eq.(8) is transformed to 

$$
\displaystyle\frac{d^2 G}{d \zeta^2}=0.
\eqno{(15)} 
$$

The solution is $G(\zeta) = C_1\cdot \zeta+C_2,$ where $C_1, C_2$ are integration constants. From (14) we have $\mu \equiv \mu_0.$ In other words, it is the Schwarzschild interior ball model. 

{\bf 2.} If we include only $a_0 \neq 0,$ then $\Omega^2 \equiv \Omega_0^2=const$ and 
$$
G \propto cos(\Omega_0\zeta+\varphi_0).
\eqno{(16)} 
$$

In this case 
$$
\mu = \mu_0 -3a_0 y =\mu_0 -3a_0 x^2,
\eqno{(17)} 
$$
\noindent
i.e. we have a parabolic distribution of the mass density (see \cite{1}). 

\vspace{.1cm}
We must say that in these cases the approximate solutions coincide with the exact well-known solutions of the Einstein equations.

{\bf 3.} Now we take $\zeta$ approximately as $\zeta \approx y/2$ ($\zeta(0)=0$), because there is the difficulty in determining the variable $\zeta$ via $y.$ Here we have an approximate estimate because the variable is $y << 1.$ In this case the origins of the two variables $y$ and $\zeta$ are glued.

The further approximation will be 
$$ 
\Omega^2 = a_0 + a_1 y \approx a_0 + 2a_1\zeta,
\eqno{(18)} 
$$
\noindent
and Eq.(7) can be written as 
$$
\displaystyle\frac{d^2 G}{d \zeta^2} + (a_0+2a_1 \zeta) G = 0,
\eqno{(19)} 
$$
\noindent
while the mass density is 
$$
\mu \approx \mu_0-6a_0 \zeta - 10a_1\zeta^2 
$$
$$
\approx \mu_0-3a_0 x^2 - (5/2)a_1 x^4.
\eqno{(20)} 
$$

The solution of Eq.(19) is
$$
G =C_1 AiryAi{\left(\displaystyle\frac{a_0+2a_1\zeta}{(2a_1)^{2/3}}\right)}+
$$
$$
+ C_2 AiryBi{\left(\displaystyle\frac{a_0+2a_1\zeta}{(2a_1)^{2/3}}\right)},
\eqno{(21)} 
$$
\noindent
where $AiryAi$ and $AiryBi$ are the Airy special functions, and $C_1,C_2$ are constants. 

{\bf 4.} The following real solution with 
$$
\Omega^2 \approx a_0 + 2a_1\zeta - 4a_2\zeta^2
\eqno{(22)} 
$$
\noindent
and
\vspace{-.3cm}
$$
\mu \approx \mu_0-6a_0 \zeta -10a_1\zeta^2+(56/3)a_2\zeta^3 
$$
$$
\approx \mu_0-3a_0 x^2 - (5/2)a_1 x^4 + (7/3)a_2 x^6
\eqno{(23)} 
$$
\noindent
can be written as the linear combination of hypergeometric functions:
$$
G=[C_1 hypergeom(\alpha_1, \beta_1, \gamma(\zeta))+
$$
$$
C_2 hypergeom(\alpha_2,\beta_2,\gamma(\zeta))(4a_2\zeta-a_1)] 
$$
$$
 \times exp(\delta(\zeta),
\eqno{(24)} 
$$
\noindent
where $C_1, C_2$ are constants,
$$
\alpha_1=-(a_1/4+a_0a_2-2a_2^{3/2})/(8a_2^{3/2});
$$ 
$$\beta_1=1/2;\qquad \alpha_2=\alpha_1+1/2;
$$ 
$$
\beta_2=\beta_1+1;\quad \gamma(\zeta)=(4a_2\zeta - a_1)^2/(8a_2^{3/2});
$$
$$
\delta(\zeta)=\zeta(a_1/2-a_2\zeta)/\sqrt{a_2}.
$$ 

{\bf 5.} One more real solution can be found for 
$$
\Omega^2 \approx a_0 + 2a_1\zeta - 4a_2\zeta^2 - 16 a_4\zeta^4
\eqno{(25)} 
$$
$$
\mu \approx \mu_0-6a_0\zeta -10a_1\zeta^2+(56/3)a_2\zeta^3
$$
$$
-(352/5)a_4\zeta^5\approx \mu_0-3a_0 x^2 - (5/2)a_1 x^4 
$$
$$
+(7/3)a_2 x^6 +(11/5)a_4 x^{10}. 
$$

This solution is written as a linear combination of HeunT functions

$$
G=C_1 HeunT(\alpha,-\beta,\gamma,-b\cdot\zeta)\cdot exp(\psi(\zeta))+
$$

$$ 
+C_2 HeunT(\alpha,\beta,\gamma,b\cdot\zeta)\cdot exp(-\psi(\zeta)),
\eqno{(26)} 
$$
\noindent
where $C_1, C_2$ are constants,
$$
\alpha = 3^{2/3} (4a_4a_0+a_2^2)/(16a_4^{4/3});\quad \beta = 3a_1/(4\sqrt{a_4}),
$$ 
$$
\gamma = 3^{1/3}a_2/(2a_4^{2/3}), b=(2/3) 3^{2/3}a_4^{1/6}, 
$$
$$
\psi(\zeta) = \zeta\cdot (3a_2+8a_4\zeta^2)/(6\sqrt{a_4}). 
$$

The pressure can be found from the equation Eq.(6) and the metric function $L$ from (7).

\section{Summary}

In conclusion, we must note that an approach to the construction of the ball static models is demonstrated in this paper. This approach is based on the reduction 
of gravitational equations to the oscillatory-type equation and the using the expansion in the power series the function which plays a role of the frequency.  Main 
difficulty is to find the new variable through the dimensionless radial variable. Five examples are considered. Two from them are the exact solutions of the Einstein equations for a fluid ball. The third, fourth and fifth examples are connected with the special Airy functions,  with the hypergeometric functions and the HeunT functions.


\begin{thebibliography}{99}

\bibitem{1}
A.M.Baranov,  Vestnik of Krasnoyarsk State University (Phys. \& Math. Sci.), No.1, 5-12 (2002)(in Russian)).

\end{thebibliography}
\end{document}